# Variant assumptions made in deriving equilibrium solutions to "A model of cardiovascular disease giving a plausible mechanism for the effect of fractionated low-dose ionizing radiation exposure" (*PloS Comput Biol* 2009 5(10) e1000539)

Mark P. Little[a, b], Anna Gola[a], Ioanna Tzoulaki[a], Wendy Vandoolaeghe[a]

[a]Department of Epidemiology and Public Health, Imperial College Faculty of Medicine, Norfolk Place, London W2 1PG, UK
[b]To whom all correspondence should be addressed at: tel +44 (0)20 7594 3312; Fax +44 (0)20 7402 2150; Email mark.little@imperial.ac.uk

**Abstract**

The paper of Little *et al.* (*PloS Comput Biol* 2009 5(10) e1000539) outlined a system of reaction-diffusion equations that were used to describe induction of atherosclerotic disease. These were solved by considering an equilibrium solution and small perturbations around this equilibrium. Here we consider slight variant sets of assumptions that could be used to derive equilibrium solutions. Although these result in equilibrium intimal bound lipid and macrophage concentrations being non-zero (unlike the earlier paper), in general they do not imply any change in the numerical results relating to monocyte chemo-attractant protein-1 (MCP-1) presented in that paper.

Equation (7) in Little *et al.* (2009) indicates that:

$$\frac{\partial \eta}{\partial t} + \nabla \cdot [\eta \chi_M(C) \nabla C] = \rho_{in}(\eta/M) M L_0 - d_M(\eta/M)\eta + \nabla \cdot [(\eta/M) D_M \nabla M] \qquad (1)$$

where $C$ is the chemo-attractant (monocyte chemo-attractant protein 1 (MCP-1)) concentration, $M$ is the macrophage concentration and $\eta$ is the bound lipid concentration. $\chi_M(C)$ is the chemotactic factor (assumed constant) associated with macrophages; the mechanism for chemotaxis is similar to that of Keller and Segel (1971a, b). $D_M$ is the rate of diffusion of the macrophages. In equilibrium let $\eta_{eq}, M_{eq}, C_{eq}$, etc. be the values of the various quantities, and let $\eta_\Delta, M_\Delta, C_\Delta$ be the differences from these equilibrium values after perturbation – so that, for example, $C = C_{eq} + C_\Delta$, and similarly for the other species. As in Little *et al.* (2009) we must have that in equilibrium:

$$\rho_{CE} E_{1,eq} + \rho_{CM} M_{eq} + \rho_{CT} T_{eq} = d_{CM} M_{eq} C_{eq} + d_{CT} T_{eq} C_{eq} + d_{Cm} m_{eq} C_{eq} \qquad (2)$$

$$\rho_{PT} T_{eq} = 0 \qquad (3)$$

$$\mu m_{eq} P_{eq} = \rho_M m_{eq} P_{eq} \qquad (4)$$

$$\rho_M m_{eq} P_{eq} = d_M(\eta_{eq}/M_{eq}) M_{eq} \qquad (5)$$

$$\rho_{in}(\eta_{eq}/M_{eq}) M_{eq} L_0 = d_M(\eta_{eq}/M_{eq}) \eta_{eq} \qquad (6)$$

$$d_T T_{eq} = 0 \qquad (7)$$

$$d_M(\eta_{eq}/M_{eq})[\eta_{eq} + d_{MM} M_{eq}] + d_T d_{TT} T_{eq} = 0 \qquad (8)$$

All variable and parameter definitions are as in Little *et al.* (2009). Assuming $\rho_{PT} \neq 0$ then by (3):

$$T_{eq} = 0 \qquad (9)$$

If $\mu \neq \rho_M$ then by (4):

$$P_{eq} = 0 \text{ or } m_{eq} = 0 \qquad (10)$$

If $d_M(\eta_{eq}/M_{eq}) \neq 0$ and $\mu \neq \rho_M$ then by (4) and (5):

$$M_{eq} = 0 \qquad (11)$$

If $d_M(\eta_{eq}/M_{eq}) \neq 0$ and $\mu = \rho_M$ (so that $M_{eq}$ is not necessarily 0) by (6) we have that:

$$\eta_{eq}/M_{eq} = \rho_{in}(\eta_{eq}/M_{eq}) L_0 / d_M(\eta_{eq}/M_{eq}) \qquad (12)$$

Under the assumption that $M_{eq} \neq 0$ equation (12) can be iteratively solved (using the parameter values of Little *et al.* (2009)) to yield:

$$\eta_{eq}/M_{eq} = 4.63 \times 10^{-17} \text{ M cell}^{-1} \qquad (13)$$

and at this value the functions $\rho_{in}(\eta_{eq}/M_{eq}), d_M(\eta_{eq}/M_{eq})$ can be evaluated (using the parameter values of Little *et al.* (2009)) to give:

$$\rho_{in,eq} = \rho_{in}(\eta_{eq}/M_{eq}) = 5.34 \times 10^{-10} \text{ cell}^{-1} \text{ ml s}^{-1} \ (\approx \rho_{in,0}) \qquad (14)$$

$$d_{M,eq} = d_M(\eta_{eq}/M_{eq}) = 3.55 \times 10^{-3} \text{ s}^{-1} \qquad (15)$$

If we perform the obvious linearisations in equation (1), and make use of the parametric forms for the macrophage bound-lipid ingestion rate:

$$\rho_{in}(\eta/M) = \rho_{in,high} + [\rho_{in,0} - \rho_{in,high}] \exp[-R_3 \eta/M] \qquad (16)$$

and the macrophage mortality rate:

$$d_M(\eta/M) = d_{M,0} + R_2 \eta/M \qquad (17)$$

assumed by Little *et al.* (2009) we obtain:

$$\frac{\partial \eta_\Delta}{\partial t} + \chi_M(C)\left[\nabla \eta_\Delta \cdot \nabla C_\Delta + \eta_{eq}\nabla^2 C_\Delta\right] = \rho_{in,eq}M_\Delta L_0$$
$$+L_0\rho_{in,high}R_3 \exp\left[-R_3\eta_{eq}/M_{eq}\right][\eta_\Delta - \eta_{eq}M_\Delta/M_{eq}] \quad (18)$$
$$-d_{M,eq}\eta_\Delta - R_2[\eta_{eq}\eta_\Delta/M_{eq} - \eta_{eq}^2 M_\Delta/M_{eq}^2]$$
$$+D_M\left[(1/M_{eq})\nabla \eta_\Delta \cdot \nabla M_\Delta - (\eta_{eq}/M_{eq}^2)|\nabla M_\Delta|^2 + (\eta_{eq}/M_{eq})\nabla^2 M_\Delta\right]$$

If we neglect second and higher order terms in $\eta_\Delta, M_\Delta, C_\Delta$:

$$\frac{\partial \eta_\Delta}{\partial t} + \chi_M(C)\eta_{eq}\nabla^2 C_\Delta \approx \rho_{in,eq}M_\Delta L_0 + L_0\rho_{in,high}R_3 \exp\left[-R_3\eta_{eq}/M_{eq}\right][\eta_\Delta - \eta_{eq}M_\Delta/M_{eq}]$$
$$-d_{M,eq}\eta_\Delta - R_2[\eta_{eq}\eta_\Delta/M_{eq} - \eta_{eq}^2 M_\Delta/M_{eq}^2] + D_M(\eta_{eq}/M_{eq})\nabla^2 M_\Delta \quad (19)$$

[Parenthetically, we notice that in the limit assumed by Little *et al.* (2009), of $\lim_{n\to\infty}\eta_{eq,n} = 0$, $\lim_{n\to\infty} M_{eq,n} = 0$, $\lim_{n\to\infty}\eta_{eq,n}/M_{eq,n} = 0$, with each set of $_n(\eta_{\Delta,n}, M_{\Delta,n}, \eta_{eq,n}, M_{eq,n})$ satisfying:

$$\frac{\partial \eta_{\Delta,n}}{\partial t} + \chi_M(C)\eta_{eq,n}\nabla^2 C_\Delta = \rho_{in}M_{\Delta,n}L_0 + L_0\rho_{in,high}R_3 \exp\left[-R_3\eta_{eq,n}/M_{eq,n}\right][\eta_{\Delta,n} - \eta_{eq,n}M_{\Delta,n}/M_{eq,n}]$$
$$-d_M\eta_{\Delta,n} - R_2[\eta_{eq,n}\eta_{\Delta,n}/M_{eq,n} - \eta_{eq,n}^2 M_{\Delta,n}/M_{eq,n}^2] + D_M(\eta_{eq,n}/M_{eq,n})\nabla^2 M_{\Delta,n}$$
$$(19')$$

then by the Arzelà-Ascoli theorem (Kelley 1975, chapter 7), and by considering a subsequence if necessary, $\eta_\Delta \equiv \lim_{n\to\infty}\eta_{\Delta,n}$ and $M_\Delta \equiv \lim_{n\to\infty} M_{\Delta,n}$ exist and satisfy expression (19) in which we replace $\eta_{eq}$, $M_{eq}$ and $\eta_{eq}/M_{eq}$ by 0 (in particular $d_{M,eq}, \rho_{in,eq}$ by $d_M(0), \rho_{in}(0)$).

Therefore, in this limit expression (19) reduces to:
$$\frac{\partial \eta_\Delta}{\partial t} \approx \rho_{in}(0)M_\Delta L_0 - [d_M(0) - L_0\rho_{in,high}R_3]\eta_\Delta \quad (20)$$

For the values of the parameters $d_M(0), L_0, \rho_{in,high}, R_3$ given in Little *et al.* (2009) we have that $d_M(0) = 9.3 \times 10^{-7}\,\text{s}^{-1}$ whereas $L_0\rho_{in,high}R_3 = 1.7 \times 10^{-17}\,\text{s}^{-1} \ll d_M(0)$. Therefore, to a good approximation (20) reduces to:

$$\frac{\partial \eta_\Delta}{\partial t} \approx \rho_{in}(0)M_\Delta L_0 - d_M(0)\eta_\Delta \quad (21)$$

in other words, equation (37) of Little *et al.* (2009).]

Integrating over the intima, and using Green's first identity and the fact that macrophage and chemo-attractant flux over the boundary is generally zero:

$$\frac{d}{dt}\int_{\Omega_I} \eta_\Delta dx = \rho_{in,eq}L_0 \int_{\Omega_I} M_\Delta dx$$
$$+L_0\rho_{in,high}R_3 \exp\left[-R_3\eta_{eq}/M_{eq}\right]\left[\int_{\Omega_I}\eta_\Delta dx - (\eta_{eq}/M_{eq})\int_{\Omega_I} M_\Delta dx\right] \quad (22)$$
$$-d_{M,eq}\int_{\Omega_I}\eta_\Delta dx - R_2(\eta_{eq}/M_{eq})\int_{\Omega_I}\eta_\Delta dx + R_2(\eta_{eq}/M_{eq})^2\int_{\Omega_I} M_\Delta dx$$

$$=\left[\begin{array}{l}\rho_{in,eq}L_0 - L_0\rho_{in,high}R_3(\eta_{eq}/M_{eq})\exp\left[-R_3\eta_{eq}/M_{eq}\right]\\ +R_2(\eta_{eq}/M_{eq})^2\end{array}\right]\int_{\Omega_I} M_\Delta dx$$
$$+\left[L_0\rho_{in,high}R_3 \exp\left[-R_3\eta_{eq}/M_{eq}\right] - d_{M,eq} - R_2(\eta_{eq}/M_{eq})\right]\int_{\Omega_I}\eta_\Delta dx$$

This has the same functional form as equation (46) of Little *et al.* (2009) although the constants are slightly different. As can be seen from the form of equations (42)-(48) of Little *et al.* (2009) this change makes no difference to the numerical estimates of averaged MCP-1 etc. given in the paper.

**Acknowledgments**
This work has been partially funded by the European Commission under contract FP6-036465 (NOTE).